\documentclass[prd, 10pt,aps,nofootinbib, floatfix, notitlepage,twocolumn,longbibliography,superscriptaddress]{revtex4-2}

\usepackage{amsmath,amsfonts,amssymb}
\usepackage[colorlinks=true,citecolor=blue,urlcolor=blue]{hyperref}
\usepackage{graphicx}
\usepackage{mathrsfs}
\usepackage{xcolor}
\usepackage{mathtools}

\def\parfrac#1#2{{\left(\frac{#1}{#2}\right) }}

\newcommand{\crit}{\text{crit}}

\makeatletter 
    
\renewcommand\onecolumngrid{
\do@columngrid{one}{\@ne}%
\def\set@footnotewidth{\onecolumngrid}
\def\footnoterule{\kern-6pt\hrule width 1.5in\kern6pt}%
}

\renewcommand\twocolumngrid{
        \def\footnoterule{
        \dimen@\skip\footins\divide\dimen@\thr@@
        \kern-\dimen@\hrule width.5in\kern\dimen@}
        \do@columngrid{mlt}{\tw@}
}%

\makeatother    

\begin{document}
\title{Lighten up Primordial Black Holes in the Galaxy with the QCD axion: \\ Signals at the LOFAR Telescope}

\author{Ricardo Z.~Ferreira}
\email{rzferreira@uc.pt}
\affiliation{CFisUC, Department of Physics, University of Coimbra, P-3004 - 516 Coimbra, Portugal}
\author{Ángel Gil Muyor}
\email{agil@ifae.es}
\affiliation{Institut de Física d’Altes Energies (IFAE) and Barcelona Institute of Science and Technology (BIST), Universitat Autònoma de Barcelona, 08193 Bellaterra, Barcelona, Spain}

\date{\today}
\begin{abstract}

\noindent In this work, we study the luminosity that results from the conversion of QCD axion particles into photons in the magnetic field of the plasma accreting onto black holes (BHs). 
For the luminosities to be large two conditions need to be met: i) there are large numbers of axions in the PBH surroundings as a result of the so-called superradiant instability; ii) there exists a point inside the accreting region where the plasma and axion masses are similar and there is resonant axion-photon conversion. 
For BHs accreting from the interstellar medium in our galaxy, the above conditions require the black hole to have subsolar masses and we are therefore led to consider a population of primordial black holes (PBHs).
In the conservative window, where we stay within the non-relativistic behavior of the plasma and neglect the possibility of non-linear enhancement via magnetic stimulation, the typical frequencies of the emitted photons lie on the low-radio band. We thus study the prospects for detection using the LOFAR telescope, assuming the PBH abundance to be close to the maximal allowed by observations. We find that for PBH and  QCD axion with masses in the range $10^{-5}-10^{-4}\, M_\odot$ and $4 \times 10^{-8}$ and $4 \times 10^{-7}$~eV,  respectively, the flux density emitted by the closest PBH, assuming it accretes from the warm ionized medium, can be detected at the LOFAR telescope. Coincidently, the PBH mass range coincides with the range that would explain the microlensing events found in OGLE. This might further motivate a dedicated search of these signals in the LOFAR data and other radio telescopes.
\end{abstract}
\maketitle

\section{Introduction}

Compact astrophysical objects such as core-collapse supernovae or neutron stars can be efficient QCD axion \textit{factories} and therefore they provide important input in the quest to detect the QCD axion. 
Indeed, SN1987A explosion still provides the strongest upper bounds on the QCD axion mass \cite{Raffelt:2006cw,Carenza:2019pxu} and neutron stars have also been shown to provide interesting input in this direction \cite{Buschmann:2021juv}. 
But compact objects can also be efficient QCD axion \textit{converters}, rather than factories, when they host large magnetic fields, such as those present in magnetars or white dwarfs. The presence of a plasma mass allows for a resonant conversion of non-relativistic axions into radio photons, and recent studies have already been able to place constraints using radio telescope data \cite{Hook_2018,Millar:2021gzs,Witte:2021arp,Battye:2019aco,Foster:2022fxn,Dessert:2022yqq,caputo2023pulsar,Safdi_2019,Huang_2018,Leroy_2020}.

What about black holes (BHs)? What is peculiar about the BH case, when compared to the other compact objects, is that the QCD axion can be dynamically generated in very large numbers, in a cloud surrounding the BH, if the so-called superradiant condition is satisfied \cite{Teukolsky_1972,Cardoso_2004}. The cloud can be very massive and store up to $\sim 10\%$ of the black hole mass \cite{East:2017ovw,Herdeiro:2021znw}, and this purely gravitational phenomenon has already been used to exclude the QCD axion for the lightest masses \cite{Arvanitaki_2015,Arvanitaki_2017,Baryakhtar:2020gao,Mehta:2020kwu}. 

However, this might not be the only way to probe the QCD axion with BHs. In contrast with other axion-like particles, the QCD axion necessarily couples photons with a strength that is mostly fixed by the axion mass.
Therefore, if some mechanism is able to convert even a small percentage of the axions in the superradiant cloud into photons in the black hole surroundings, this can open new observational probes. In that sense, one possibility that has been considered is that the axions in the cloud decay to two photons, and stimulation effects can then result in fast radio bursts \cite{Rosa_2018,Ikeda_2019}.
The possibility of axion-photon conversion on the magnetic fields surrounding the cloud has also been discussed in previous literature \cite{Arvanitaki_2010,Arvanitaki_2011}. 

In this work, we will explore how the the accretion of plasma and magnetic field from the interstellar medium (ISM) onto the black hole can affect this picture and argue that the situation becomes closer to the resonant conversion that happens, for example, in magnetars.
Axions can decay to two photons as long as the axion mass $m_a$ is larger than twice the plasma mass $\omega_p$. However, the presence of the magnetic field opens up a new channel, axion-to-photon conversion, that is typically more efficient than the decay to two photons and is, in fact, the only process kinematically allowed when $\omega_p> m_a/2$.  
Moreover, the plasma density and the magnetic field increase towards the BH horizon at a pace that depends on the accretion flow. Then, as long as the plasma mass of the medium is smaller than the QCD axion mass outside the accreting region, there can be a critical radius inside the accreting region where the QCD axion and the plasma mass are close to each other thus allowing resonant conversion between the two. If the critical radius happens to be close to the position of the superradiant cloud, where the axion densities are higher, it is then likely that many QCD axions will convert into photons. 
The ISM magnetic fields are much weaker than those observed in magnetars, but that can be largely compensated by the enormous number of axions in the cloud thus boosting the luminosity of the process. Moreover, contrarily to the case of neutron stars or white dwarfs, here the axions that convert into photons are dynamically generated by the superradiant instability, so the mechanism does not rely on the assumption of an initial axion profile.

However, when restricting to BHs in our galaxy, imposing that the critical radius occurs close to the superradiant cloud typically requires the BHs to have subsolar masses and we are therefore led to assume a population of primordial black holes (PBHs). We will then show that in the most conservative region, where we neglect the possibility of non-linear enhancement via magnetic stimulation and where the electrons are non-relativistic at the conversion point, the photons emitted from the resonance are on the low-radio band and can readily be searched for with the LOFAR radio telescope \cite{vanHaarlem:2013dsa}. Coincidently, the region probed by LOFAR corresponds to PBH with masses in the range where there have been recent hints from microlensing events \cite{Niikura:2019kqi}. 

The manuscript is organized as follows.
In section \ref{sec:Superradiance} we provide a brief summary of the superradiant phenomena and give the basic properties of the cloud and of the QCD axion. Section \ref{sec:Interstellar medium} describes the typical properties of the interstellar medium in our galaxy in terms of electron density, temperature and magnetic fields. 
Section \ref{sec:Accretion models} concerns the Bondi accretion onto black holes and the associated accreting profiles. The main mechanism of this work, the resonant conversion, is described in section \ref{sec:Resonant conversion wo gravitational potential}. In section \ref{sec:Fate of the conversion}, we discuss the observational prospects for detection with LOFAR. Finally in section \ref{sec:Outlook} we conclude and give some outlook for the future. Most of the technical details are left for the appendices. In appendix \ref{app:Media} we give the typical values of the properties of the interstellar medium that we use in this work. In appendix \ref{app: Method of variation of parameters} and \ref{app: Solution of the homogeneous equation} we explain how we use the method of variation of parameters to find a solution for the equation of motion of the photon around the superradiant cloud, and find, in appendix \ref{app:Analytics}, a simplified expression for the luminosity of the system. Finally, in appendix \ref{app:Stimulation emission} we discuss the phenomena of magnetic stimulation and discuss the region of parameters where we expect that to happen.

\section{Superradiance and the QCD axion}
\label{sec:Superradiance}

Via the superradiant instability, it is possible to convert a significant part of the BH spin into dense QCD axions clouds \cite{Arvanitaki_2011}. 
The axions in these clouds are in quasi-bound states similar to the hydrogen atom, labeled by  $n$, $l$, and $m$ and with energies $\omega_{nlm} \simeq m_a(1-\alpha^2/2n^2)$, where
\begin{equation}
\alpha=G M m_a 
\label{eq:alpha}
\end{equation}
is the gravitational coupling, analogous of the fine-structure constant, and $m_a$ is the QCD axion mass. 
For the superradiant phenomena to be possible, the condition $m \Omega-\omega_{nlm}>0$ has to be satisfied, where $\Omega$ is the BH angular velocity. This condition enforces the restriction $\alpha<\tilde{a} m/2$ where $|\tilde{a}|<1$ is the dimensionless spin parameter of the BH.

Among the different superradiant states, the 2p state with $nlm=211$, is the fastest growing one with rate \cite{Detweiler_1980}
\begin{equation} \label{eq: Superradiant rate}
	\Gamma_{211} \simeq \frac{\alpha^8 (\tilde{a}-4 \alpha)}{24}  m_a  \,.
\end{equation}
We will focus on the axions that populate this state as these exist in larger numbers than those in other states. 

The axion field in this level $\bar{a}$ has a profile given by
\begin{eqnarray}
\bar{a}(t, \vec{r}) =\sqrt{\frac{N}{2m_a}}\left[ e^{-i \omega_{211} t} R_{21}(r) Y_{11}(\hat{r})+h.c.\right]
\label{eq: non-rel red}
\end{eqnarray}
where $N$ is the number of axions in the cloud, $Y_{11}(\theta,\phi) $ is the $l=m=1$, spherical harmonic and
\begin{eqnarray}
    R_{21}(r)= \parfrac{1}{24a_0^3}^{1/2}  \frac{r}{a_0} \, e^{-r/(2a_0)}
\end{eqnarray}
is the radial bound state solution of the  $211$ level with $a_0=1/(m_a \alpha)=G M/\alpha^2$ the gravitational Bohr radius of the system, so that the average distance from the BH is 
\begin{eqnarray}
 \label{eq:Cloud radius}
r_\text{cl} \equiv \frac{\int r^3 R_{21}^2\, dr}{\int r^2 R_{21}^2 \, dr} = 5a_0 \,.
\end{eqnarray}

Regarding the spin, each axion in the 2p state has $m=1$ and so the angular momentum in the cloud is equal to the total number of axions $J_\text{cl}=N$. Therefore, once most of the BH angular momentum $ J_\text{BH} =\Tilde{a}GM_\text{BH}^2$ is drained to the cloud $J_\text{cl}\simeq J_\text{BH}$, the cloud would get a mass   
\begin{equation}
M_\text{cl} = N m_a  = \tilde{a}\alpha M_\text{BH}
    \label{eq: mass of the cloud} \, .
\end{equation}

In the following sections, we will restrict to situations where the cloud had time to grow substantially until the present time. Therefore, we require the superradiance rate to be faster than the age of the universe which, jointly with the superradiant condition $\alpha<\tilde{a}/2$, sets the following upper and lower bounds on the QCD axion mass~\footnote{In terms of the axion decay constant $f_a$, the range is \begin{eqnarray} 
	\frac{9.4 \times 10^{18}}{\tilde{a}^{-1/9}} \parfrac{M}{ M_\odot}^{8/9}\gtrsim \,f_a/\text{GeV} \,
\gtrsim  \frac{8.5\times 10^{16}}{\tilde{a}}  \,\parfrac{M}{ M_\odot} \,.
\end{eqnarray}}
\begin{eqnarray} \label{eq: maximal and minimal ma}
	\frac{5.3 \times 10^{-4} }{\tilde{a}^{1/9}} \parfrac{ M_\odot}{M}^{8/9}\lesssim \,\frac{m_a}{10^{-9}\text{ eV}} \,
\lesssim  0.067 \,\tilde{a} \,\parfrac{ M_\odot}{M} \,.\ 
\end{eqnarray}

Current observational bounds restrict the QCD axion mass to lie roughly in the interval $10^{-11}-10^{-1}$ eV \cite{DiLuzio:2020wdo}. In particular, the absence of a superradiance cloud around the observed rotating stellar mass black holes is what sets the lower bound \cite{Baryakhtar:2020gao}.
We are therefore led to consider QCD axion clouds that form around primordial black holes (PBH) that we will assume to exist with a fraction $f_\text{PBH}$ of the total dark matter abundance. Moreover, we will be interested in the coupling between the QCD axion and the photon which has the form
\begin{equation}
\mathcal{L}\supset \frac{g_{a\gamma}}{4} aF_{\mu\nu} \tilde{F}^{\mu\nu} 
\label{eq:photonCoupling}
\end{equation}
where $F$ is the electromagnetic field strength tensor, $\tilde{F}$ is its dual and $g_{a\gamma}$ is the axion-photon coupling constant  which, in the QCD axion case, is directly related to its mass via the relation
\begin{equation}
g_{a\gamma}=2\times 10^{-19}  c_\gamma \parfrac{m_a}{ 10^{-9}\text{eV}}  \, \text{GeV}^{-1} \,,
    \label{eq: axion photon coupling}
\end{equation}
where $c_\gamma$ is a model-dependent parameter of order one \cite{DiLuzio:2020wdo}. In this work, we will use the conservative value of $c_\gamma=1$. 

\section{Accretion from the Interstellar medium}
\label{sec:Interstellar medium and accretion}

The PBH hosting the QCD axion cloud does not live in isolation. In this section, we discuss the properties of the astrophysical environment surrounding the system. We will focus on PBHs that accrete from the interstellar medium (ISM) in our galaxy and so start by discussing a few relevant properties of the ISM that play a major role in the axion-photon conversion: the photon plasma mass $\omega_p$, the magnetic field $B$ and the sound speed $c_s$ (that we summarize in Table \ref{tab: Components of the interstellar medium}). We then proceed with a discussion of the accreting dynamics onto the PBH and the resulting profiles.

\subsection{Interstellar medium}
\label{sec:Interstellar medium}

Starting from the photon plasma mass, in the non-degenerate and non-relativistic limit, it is given by
\begin{eqnarray} \label{eq:plasma frequency definition}
	\omega_p&=& \parfrac{4\pi \alpha_\text{EM} n_e}{m_e}^{1/2}\nonumber\\ &=& 1.2 \times 10^{-11} \parfrac{n_e}{0.1 \text{cm}^{-3}}^{1/2}\, \text{eV},
\end{eqnarray}
where $m_e$ is the electron mass, $\alpha_\text{EM}$ is the electromagnetic coupling constant and $n_e$ is the electron density that we normalized to the typical value in the warm ionized medium. The electron density is related to the medium density $n_\infty$ by the ionization fraction $x_e$. The plasma mass can go down by one order of magnitude in the more rarified hot ionized medium but it can also be larger in denser components of the ISM.  
In Table \ref{tab: Components of the interstellar medium} we show typical values of these quantities.
For the conversion to take place, the frequency of the axions in the cloud $\omega_a=\omega_{211} \approx m_a$  needs to be larger than the photon plasma mass,  
\begin{eqnarray} \label{eq: mass constraint}
     \omega_a  \geq \omega_p \, . 
\end{eqnarray}
The accretion of the plasma onto the black hole will further increase its local density and thus place stronger constraints on the possible axion masses that can be converted into photons, as we will study in detail in the following sections. Smaller ISM densities allow a wider range of axions to convert into photons; for that reason, and because of the large filling factors that ensure that there are many PBH in such environments \cite{Draine_2010}, we will mostly focus on the warm and hot ionized components of our galaxy.~\footnote{In the extragalactic space, one expects smaller densities, that can go down to $10^{-7}\text{cm}^{-3}$ \cite{Dalton:2021egz}, and would allow the conversion of lighter QCD axion. However, here we focus on the conversion within our galaxy and leave the extragalactic case for future work.}

Another important property for the axion-photon conversion is the magnetic field. In the ISM their typical strength is of order $\mu G$ in the hot ionized bubble and the warm neutral medium \cite{Ferriere_2001,Xu2019,Pelgrims2020} but larger values, of order $10 \mu$~G, have also been found in regions of Hydrogen gas and clouds  \cite{Draine_2010}. The accretion onto the PBH will further increase the local value of the magnetic field towards the PBH  and therefore the conversion can take place at values larger than the $\mu G$, as we will discuss next. 

The last relevant property is the sound speed of the medium $c_s$ that can take values from $0.6$ km/sec in dense environments such as molecular clouds, to $100$ km/sec in hot ionized medium (c.f. Table \ref{tab: Components of the interstellar medium}). The sound speed will affect the accreting properties of the PBH and therefore the conversion.  Following \cite{Luca_2020},  
 we will take the PBH to have a speed $v_\text{rel}$ relative to the medium that is smaller or comparable to $c_s$.

\subsection{Accretion onto the BH}
\label{sec:Accretion models}

\begin{figure}[t]
    \centering
\includegraphics[width=\columnwidth]{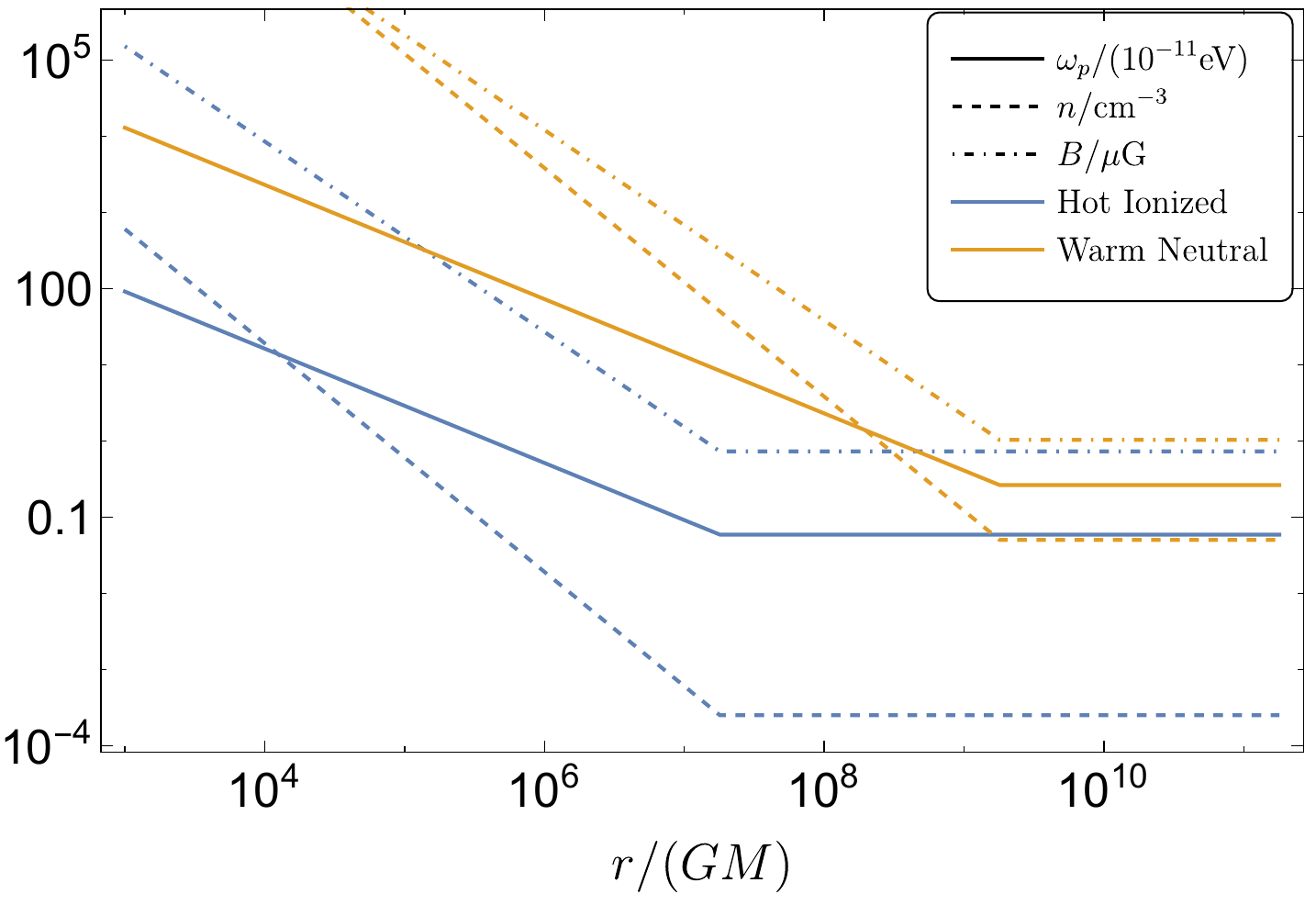}
    \caption{Profiles for the plasma density, magnetic field and medium density in the hot ionized and warm neutral media.}
    \label{fig:accreting profiles}
\end{figure}

We will assume that the PBH follows a simple model of spherical Bondi accretion \cite{Luca_2020,Chisholm:2002ba,Fender_2013}  \footnote{The assumption of Bondi accretion is justified for low-luminosity BHs accreting from low-density regions, as the ones considered in this work. }
\begin{eqnarray}
	\dot{M}= \lambda \, 	\dot{M}_b
\end{eqnarray}
where $\lambda=10^{-4}-10^{-2}$ is an efficiency parameter \cite{Fender_2013} and $ \dot{M}_b$ is the Bondi accretion rate given by  \cite{Bondi_1952,Ipser_77}
\begin{equation} 
	\dot{M}_b= \frac{\beta}{4} \pi  \, v \,r_b^2  \,n_\infty \,m_p,
	\label{eq:accretionRate}
\end{equation}
where $v= \sqrt{ v_\text{rel}^2+c_s^2}$,
\begin{eqnarray} \label{eq: Bondi radius}
	r_b= \frac{2GM}{v^2}
\end{eqnarray}
is the Bondi radius that limits the accreting region, $n_\infty$ is the medium density, $m_p$ is the proton mass and $\beta \sim {\cal O}(1)$ is a coefficient that depends on the equation of state of the medium \cite{Shapiro_1976,Ipser_77}. 

This accretion is rather inefficient and so the accreting time scales are very large. For example, for medium densities and velocities typical of the hot ionized medium, the accreting time $\tau_\text{acc}=M/\dot{M}$ is larger than the age of the universe for PBHs with (sub-)solar masses. Therefore, we can safely neglect the change in mass and spin of the PBH due to accretion. Similarly, the luminosity of the accreting flow is also very low. For sub-Eddington accretion rates as those described above the luminosity is $L= \eta \dot{M}$ where $\eta \sim \dot{M}/ L_\text{edd}$ and $L_\text{edd}$ is the Eddington luminosity \cite{Fender_2013,Manshanden_2019}, thus making these objects quite faint.

For a constant accretion rate with equipartition of kinetic and gravitational energies, the plasma velocity evolves as $v \sim \beta_v \sqrt{G M_\text{BH}/r}$, where $\beta_v$ is an order one coefficient that depends on the fluid being accreted, and the density profile can be obtained from Eq. \ref{eq:accretionRate} to be   \cite{Ipser_77}
\begin{equation}
	n(r)= \frac{\beta}{4\beta_v}  n_\infty \left(\frac{r_b}{r}\right)^{3/2} , \qquad r\lesssim r_b \, .
	\label{eq: accretion density}
\end{equation}
For $r \gtrsim r_b$ the density saturates to the medium density $n_\infty$.
As a consequence, 
the photon plasma mass in Eq. \ref{eq:plasma frequency definition} will also acquire a radial profile, since $n_e(r)=x_en(r)$. We show the profiles for the density and the photon mass in Fig. \ref{fig:accreting profiles}. The presence of a radial dependent plasma mass has the  important consequence of allowing for the presence of a critical radius $r_\text{crit}$ such that 
\begin{eqnarray}
    \omega_p(r_\text{crit})= m_a \, , \label{eq: rcrit definition}
\end{eqnarray}
i.e. where the photons in the plasma and the QCD axion have the same mass and so can resonantly convert into each other.
As explained in the next section, the conversion to photons of the axion in the cloud will be dominated by the dynamics around this radius (and not at the radius of the cloud). In terms of the model parameters, the critical radius is given by
\begin{equation}
r_\text{crit}=\left(\frac{\omega_{p,\infty}^2}{4m_a^2}\right)^{2/3} r_b
    \label{eq:rcrit}
\end{equation}
where we have used Eqs. \ref{eq:plasma frequency definition}, \ref{eq: Bondi radius} and \ref{eq: accretion density},  and defined $\omega_{p,\infty}\equiv\omega_p(r\to\infty)$. We will be interested in situations where the plasma frequency of the ISM is lower than the QCD axion mass, $\omega_{p,\infty} <m_a$, and therefore the critical radius will always be inside the accreting region.

Accretion will also enhance the magnetic field towards the PBH.
Magnetic flux conservation requires the magnetic field to grow as $B\propto r^{-2}$ towards the black hole \cite{Ioka:2016bil}. However, the magnetic field pressure cannot exceed the pressure of the accreting matter \footnote{If the magnetic field pressure is larger than that given by the equipartition of energy the accretion might enter a magnetically dominated state \cite{Blandford:1977ds,Bisnovatyi-Kogan}. In this work, we will however only consider situations where the magnetic field strength is fixed by equipartition.} and so the magnetic field profile will tend towards equipartition of  magnetic, kinetic and gravitational energies \cite{Shvartsman_1971,Beskin_2005,Ipser_1982,EHT_2019}
\begin{eqnarray}
	\frac{B^2}{2} \sim \beta_B  \frac{G M}{r} m_p n(r)\,
 \label{eq: accretion B}
\end{eqnarray}
and thus grow more slowly, as $r^{-5/4}$ (see  Fig. \ref{fig:accreting profiles}). The parameter $\beta_B$ is the ratio between the gas pressure and the magnetic pressure that we take  to be of order one as in \cite{EHT_2019}. Then, for example, for the benchmark properties of the hot ionized medium the typical strength of the magnetic field expected from equipartition is
\begin{eqnarray}
	B &=&  0.7 \sqrt{\frac{\beta_B \beta}{\beta_v}}  \,\parfrac{v}{100 \,\text{km/sec}} \nonumber \\ &\times& \parfrac{n_\infty}{10^{-3}\, \text{cm}^{-3}}^{1/2} \parfrac{r_b}{r}^{5/4} \mu \text{G} 
\end{eqnarray}
consistently with the values discussed in Section \ref{sec:Interstellar medium} outside the accretion region. The direction of the magnetic field is however harder to determine and can take different topologies \cite{EventHorizonTelescope:2021srq}.
 In section \ref{sec:Resonant conversion wo gravitational potential} we make some assumptions on the directionality and briefly discuss the dependence of the results on those assumptions.

Finally, let us discuss the evolution of the plasma temperature towards the black hole.
The expression we used for the plasma frequency in Eq. \ref{eq:plasma frequency definition} is valid as long as the electrons are non-relativistic and non-degenerate. However, in hot accretion flows electrons are expected to be relativistic close to the black hole \cite{Colpi_1984,Mahadevan_1997}.
We can estimate the radius at which  electrons become relativistic by assuming that far enough from the black hole the equipartition of energy determines the electron temperature to be \cite{EHT_2019}
\begin{eqnarray} \label{eq: plasma temperature}
T \sim \frac{G M m_p}{r}  \sim 0.9 \, m_e \parfrac{10^{3}r_s}{r}\, .
\end{eqnarray}
At radius $r\lesssim r_\text{rel} \equiv 10^3 r_s$, electrons become relativistic and the expression for the plasma density in Eq. \ref{eq:plasma frequency definition} is no longer adequate.~\footnote{The value of $r_\text{rel}$ obtained agrees with more dedicated studies of hot accretion flows \cite{Mahadevan_1997,Yuan_2014}.} 
As we will see in the next section, imposing that the axion-photon conversion occurs in a regime where the electrons are non-relativistic provides strong constraints on the parameter space.

\section{Axion-photon conversion in the  black hole vicinity}
\label{sec:Resonant conversion wo gravitational potential}

Now that we have characterized the environment surrounding the black hole, we can explore its  interactions with the axions in the cloud. 

The axion-photon coupling in Eq. \ref{eq: axion photon coupling}  allows axions to decay to two photons with energies $\omega_\gamma=m_a/2$ at a rate $\Gamma=g_{a\gamma}^2 m_a^3/(64\pi)$ and has been studied in the superradiance context \cite{Rosa_2018,Ikeda_2019,Caputo_2021,Blas_2020,Blas_2020b,Branco:2023frw}.
However, axions can also be converted into photons in the accreting magnetic field. The conversion probability per unit length typically scales as $dp/dL \propto g_{a\gamma}^2 B^2 L$ where $L$ is a length scale during which the conversion takes place and it is typically related to the magnetic field coherence length or the width of the resonance band \cite{Raffelt_1987}. For the typical values of the ISM magnetic field, $B\sim \mu  \text{G}\sim 10^{-8} \text{eV}^{2} \gg m_a^2$, the conversion process, if available, will likely dominate over the decay to two photons even if $L\sim 1/m_a$. 
Furthermore, in the region where the plasma frequency is larger than $m_a/2$ but below $m_a$ the decay to photons is kinematically forbidden and the conversion is the only process allowed.

\subsection{Toy model: Photon equation in a spherically symmetric potential}

We proceed with the explicit computation of the photon luminosity from the conversion of the axions in the cloud.

The dynamics of photons in an accreting plasma around a rotating black hole has been the subject of several recent studies \cite{Rosa:2011my,Conlon_2018,Dima:2020rzg,Cannizzaro_2021,Cannizzaro:2021zbp}. In particular, there has been an interesting discussion about the possible development of photon superradiance in the presence of a plasma mass.
We disregard here this possibility, which has been argued to be unlikely for realistic astrophysical accretions \cite{Dima:2020rzg},  but rather focus on the photon free-state solutions that will be created from the presence of an axion and magnetic field profiles. We follow the approach of \cite{Dima:2020rzg} and take the Klein-Gordon equation
\begin{eqnarray} \label{eq: Klein-Gordon}
    \left[ \Box + \omega_p^2(r) \right] \bar{A}(t,\vec{r})= S(t,\vec{r}) \, 
\end{eqnarray}
 as a toy model for the propagation in the accreting region of the component of the photon field that is parallel to the magnetic field $\bar{A}$, which is the one that interacts with the axion \cite{Raffelt_1987}. We also  added a source term $S(t,\vec{r})$ that contains the information about the magnetic field and axion cloud profiles. We will assume for simplicity that at the conversion point, the magnetic field is perpendicular to the line of sight and neglect ray-bending effects. On the other hand, we leave the angle between the BH rotation axis and the line of sight free.

As discussed before, one of the constraints we need to impose is that $r_\text{crit}>10^{3}r_s$ to ensure the non-relativistic behavior of the electrons in the plasma. This means that in the region of the parameter space fulfilling this condition, the conversion point will be far from the PBH horizon and therefore we can safely take the weak field limit of Eq. \eqref{eq: Klein-Gordon}
\cite{Detweiler_1980,Baryakhtar:2020gao}
\begin{eqnarray}  \label{eq: equation of motion}
   \left[  \partial_t^2 - \nabla^2 + \omega_p(r)^2 \left(1 - \frac{2 G M_\text{BH}}{r} + \hat K \right) \right] \bar{A}(t,\vec{r}) \nonumber \\
  = g_{a\gamma \gamma}  \partial_t \bar{a}(t, \vec{r}) B(r) &&  
\end{eqnarray}
where $\nabla$ is the gradient operator in spherical coordinates. 
The terms in $\hat{K}$ includes corrections to the Newtonian potential that are suppressed by $\alpha$ in the non-relativistic limit (see Appendix B of \cite{Baryakhtar:2020gao}). We will neglect $\hat K$ in the remainder of this section.

The magnetic field has a stationary profile, therefore, energy conservation imposes that only the mode of $\bar{A}$ with frequency $\omega_a=\omega_{211} \approx m_a$ will be excited. Moreover, following Sec. \ref{sec:Accretion models}, we  consider that the $B$ field has a radial profile $B(r,\theta,\phi)\simeq B(r)$ so that the decomposition of the source term in terms of spherical harmonics only has non-vanishing coefficients for $l=m=1$ and the created photon inherits the same angular dependence.

After decomposing the photon and the scalar field, $X=\{A,a\}$, as $\bar{X}(t,\vec{r})=X(\vec{r})e^{-im_a t}+X^*(\vec{r})e^{im_a t}$ we then find
\begin{eqnarray}
    \left[  -m_a^2 - \nabla^2 + \omega_p(r)^2  \right] A(\vec{r})   = f(\vec{r})
    \label{eq: Photon equation}
\end{eqnarray}
where we have defined $
    f(\vec{r}) \equiv -i m_a  g_{a\gamma \gamma}  a(\vec{r}) B(r)$.

In Appendices \ref{app: Method of variation of parameters} and \ref{app: Solution of the homogeneous equation} we solve in detail the differential equation \ref{eq: Photon equation} and summarize here the main findings. After decomposing the solution in spherical harmonics and applying the method of variation of parameters, the behavior of the solution for $A$ at $r\to\infty$ can be written as
\begin{equation}
    A(\Vec{r})\to Y_{11}(\hat{r})\frac{e^{i \theta(k_\infty r)}}{ik_\infty r} \int f_{11}(r')F(r')r' dr'
    \label{eq: field}
\end{equation}
where $\theta$ is a phase such that $\partial_r\theta (r\to \infty) = k_\infty$ with  $k_\infty=\sqrt{m_a^2-\omega_{p,\infty}^2}$ the photon momentum at infinity, $f_{11}(r)$  the projection of $f(\vec{r})$ over the spherical harmonic $Y_{11}(\hat{r})$, and $F(r)$ the homogeneous solution of Eq. \ref{eq: Photon equation} that we find, using WKB methods, to be given by Eq. \ref{eq: Homogeneous sol}. The factor of $k_\infty$ dividing in Eq. \ref{eq: field} arises from the modulation of the outgoing wave due to the non-constant plasma mass \cite{Hook_2018}. This factor would have been missed if we had reduced  Eq. \ref{eq: Photon equation} to a first-order differential equation.

The time-averaged luminosity in the radial direction is related to the Poynting vector $\vec{E} \times \vec{B}$ of the outgoing wave, where $\vec{E}$ is the electric field, and at infinity it is given by
\begin{eqnarray}
    \left<\frac{dP}{dA}\right>=\left<\frac{dP}{r^2d\Omega}\right>&=&\left<\partial_r \bar{A}(t,\Vec{r}) \partial_t \bar{A}(t,\Vec{r})\right>\nonumber \\ &=&2\text{Re}(im_a A(\vec{r})^*\partial_r A(\vec{r})) \, .
\end{eqnarray} 
Therefore, the average photon luminosity per unit solid angle simplifies to 
\begin{eqnarray}
    \left<\frac{d P }{dA}\right> = 2 m_a k_\infty |A(\vec{r})|^2
    \label{eq: Lum def}
\end{eqnarray}
where we have used the fact that the plasma frequency asymptotes to a small constant value at the Bondi radius, 
Using Eq. \ref{eq: field} 
we then find
\begin{eqnarray}
    \left<\frac{dP}{d\Omega}\right>&= & \left|Y_{11}(\hat{r})\right|^2 \frac{m_a^2}{k_\infty}g_{a\gamma\gamma}^2N\nonumber \\ && \times \left|\int B(r')R_{21}(r')F(r')r'dr'\right|^2
    \label{eq: app Luminosity 2} \,.
\end{eqnarray}
where $N$ is the total number of axions in the cloud given in Eq. \ref{eq: mass of the cloud}.

In appendix \ref{app: Solution of the homogeneous equation} we compute the remaining integral. We find that if  $\rho_\text{crit}\equiv m_a \, r_\text{crit}> 1$, most of the support of the integration comes from a region around the critical point, $r \sim r_\text{crit}$ with width $\Delta L \sim \rho_\text{crit}^{1/3}$, where the homogeneous solution behaves like an Airy function. 
For $r \ll r_\text{crit}$  the photon mass is bigger than the axion energy and  the conversion is kinematically blocked whereas for $r\gg r_\text{crit}$ the homogeneous solution oscillates quickly and the integral is also strongly suppressed. 

Furthermore, if the condition $4\rho_\text{crit}^{2/3}/|3-2\alpha\rho_\text{crit}|\gtrsim 1$ is verified, both the magnetic field and the radial wave function are approximately constant around the conversion window thus allowing to simplify the expression for the luminosity to: 
\begin{eqnarray}
     \left<\frac{d P }{d\Omega}\right> (\hat{r}) &=&  \frac{m_a}{2}  \tilde{n}_\text{crit}\,\frac{r_\text{crit}^2}{a_0^3} (L \,  g_{a\gamma\gamma} \, B_\text{crit})^2\nonumber
    \\
    &\simeq & 8\times 10^{9}\ \tilde{a}\parfrac{\tilde{n}_\text{crit}}{N}  \parfrac{M_\text{BH}}{10^{-2} M_\odot}^{8}\parfrac{m_a}{10^{-9}\text{ eV}}^{\frac{22}{3}}\nonumber\\
    &  \times& \parfrac{n_\infty}{10^{-3}\, \text{cm}^{-3}}^{4/3} \parfrac{100 \,\text{km/sec}}{v}^{4} 
    \, \text{W}.
    \label{eq: Luminosity}
\end{eqnarray}
where  $n = \tilde{n}/a_0^3 =N|R_{21}(r_\text{crit}) Y_{11}(\hat{r})|^2$ is the number density of axions at the critical radius, $B_\text{crit}$ is the value of the magnetic field at the critical radius and we have normalized quantities to typical values in the hot ionized medium. We have also defined $L=\sqrt{\pi/\omega_p'(r_\text{crit})}= \sqrt{4\pi \rho_\text{crit}/3}$ which is sometimes called the width of the conversion region although in this context we find a smaller conversion width of  
 $\Delta L \propto L^{2/3}$. We also verified numerically that the analytical approximation gives a good estimation of the total luminosity.

\begin{figure*}
\centering
\includegraphics[width=\columnwidth]{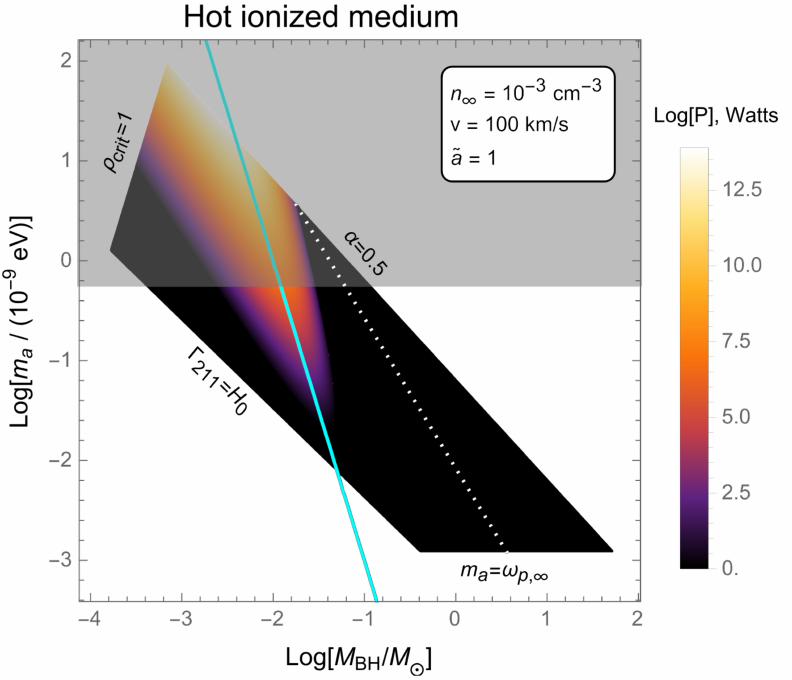}\hfill
\includegraphics[width=\columnwidth]{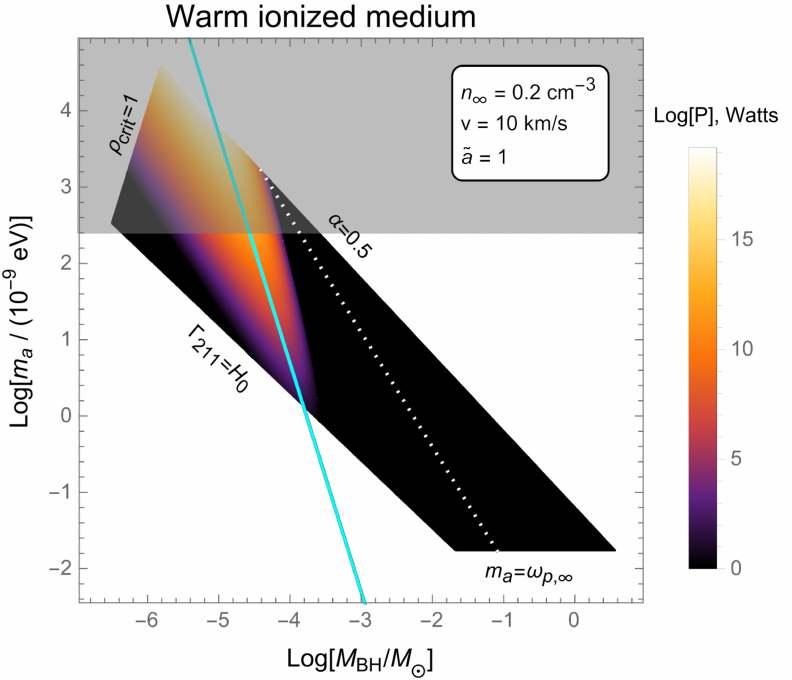}
\caption{Luminosity from the resonant conversion to photons of QCD axions in a superradiant cloud for typical values of the hot ionized medium (left plot) and warm ionized medium (right plot) and for a spin parameter $\tilde{a}$ close to one. The cyan line corresponds to when $r_\crit=r_\text{cl}$. Inside the gray region, the electrons in the plasma become relativistic at the critical radius. We have also fixed  $c_\gamma=1$ in Eq. \ref{eq: axion photon coupling}.}
\label{fig:Density plot}
\end{figure*}

\subsection{Results}

Figure \ref{fig:Density plot} shows the luminosity emitted from the axion-photon conversion, using Eq. \ref{eq: Luminosity}, for different values of the QCD axion and PBH masses and for two different media in which the PBH accretes: the hot ionized medium and warm ionized medium. 

When restricting to parameters below the gray band, we find luminosities as high as $10^7$ W for PBHs with masses around $10^{-2}M_\odot$ and QCD axion masses around $10^{-9}$ eV when the accretion happens in the hot ionized medium. For and accretion in the warm ionized medium, we find instead $10^{12}$~W for PBH masses around $10^{-5}M_\odot$ and QCD axion masses around $10^{-7}$~eV. The gray band is the region where the electrons in the accreting plasma are relativistic at the conversion radius (c.f. Eq. \ref{eq: plasma temperature}) and therefore the expression for the plasma mass in Eq. \ref{eq:plasma frequency definition} is no longer valid. We do not expect a dramatic change when entering the relativistic limit, given that the plasma mass does not change significantly, but it would require a dedicated computation that goes beyond the scope of this paper.

In Fig. \ref{fig:Density plot}, the most luminous region lies around the  cyan line which  corresponds to the cases where the conversion radius coincides with the average position of the cloud $r_\text{crit}=r_\text{cl}$ and requires
 \begin{eqnarray}
\frac{m_\text{a}}{10^{-7}\text{ eV}} \approx \parfrac{n_\infty}{10^{-3}\text{ cm}^{-3}}^{1/2}\parfrac{100\text{ km/s}}{v}^{3/2} \alpha^{3/2}  \, .
\end{eqnarray}

We have also restricted the parameter space by the constraints in  Eqs. \ref{eq: maximal and minimal ma} and Eq. \ref{eq: mass constraint} that ensure, respectively, that  superradiance happens and within the lifetime of the universe, and that the emitted photons can escape the accreting region and propagate throughout the medium. To have an analytical solution we also had to require that $\rho_\text{crit}>1$. Note that all these constraints are independent of the axion-photon coupling. 

 As we move to larger $\alpha$, we expect other effects to become important. Namely, as QCD axion mass increases both the cloud and the critical radius become closer to the PBH and gravitational corrections to Eq. \ref{eq: equation of motion} will start becoming important. On the other hand, around the cyan line, as we increase  $\alpha$ at some point the magnetic field generated from the conversion becomes larger than the original one from accretion and the system is expected to enter a magnetically stimulated regime that we discuss in Appendix \ref{app:Stimulation emission}. However, for the media considered in this work, the gravitational effects and the stimulated regime are only relevant inside the gray region and we therefore neglect them in the rest of the analysis.
 
Finally, to the right of the dotted white line, $4\rho_\text{crit}^{2/3}/|3-2\alpha\rho_\text{crit}|< 1$, the magnetic field and the axion profile start varying significantly around the resonant point and Eq. \ref{eq: Luminosity} is no longer valid. In this regime, we calculated numerically the luminosity using Eq. \ref{eq: app Luminosity 2} and verified that the luminosity is still much suppressed compared to the region around the cyan line.

\section{Detection Prospects at radio telescopes}
\label{sec:Fate of the conversion}

\begin{figure*}[t]
    \centering \includegraphics[width=\textwidth]{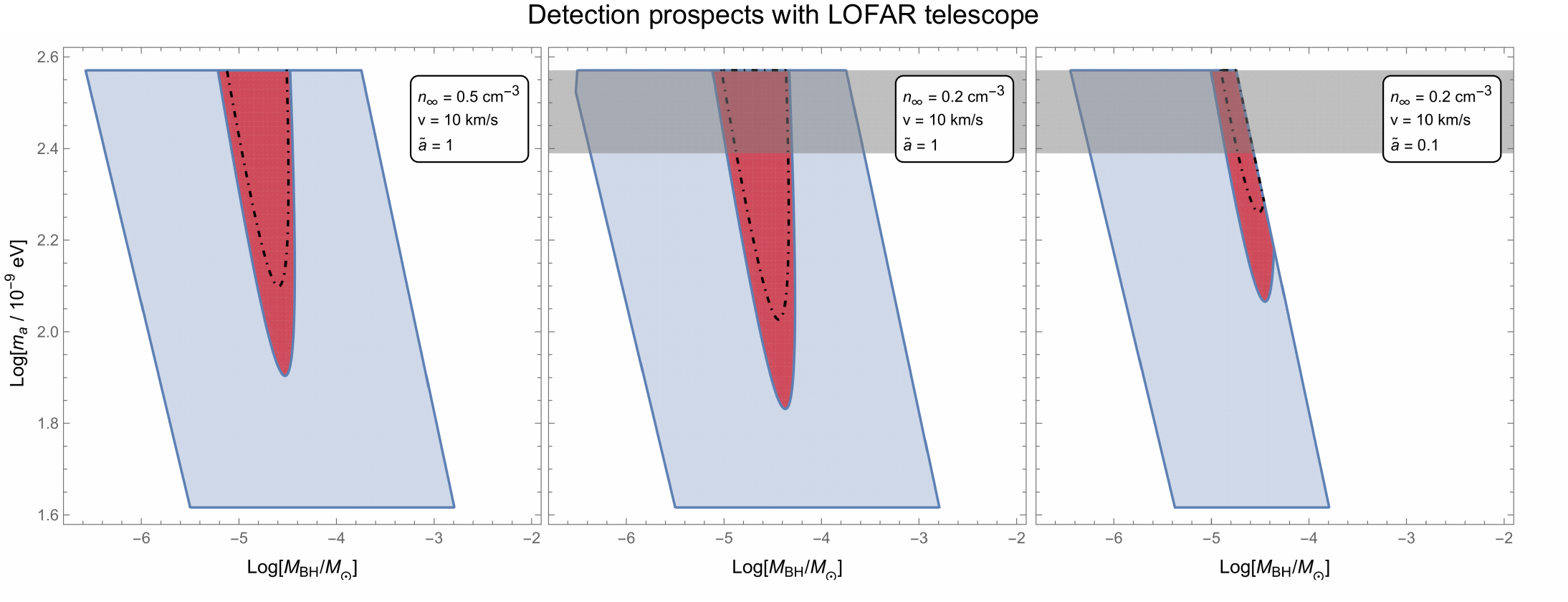}
    \caption{Parameter space (QCD axion mass vs PBH mass) where the emitted photons lie in the frequency band of LOFAR and i) satisfy the superradiant constraints as in Fig. \ref{fig:Density plot} (Blue region) and ii) can be detected at LOFAR with 8 hours of observation time (Red region). We fixed the PBH fraction to $f_\text{PBH}=0.1$, with spin parameter $\tilde{a}=0.1$ and close to $1$ and assumed the closest PBH to be accreting from the warm ionized medium with densities of $n_\infty=0.2\ \text{and } 0.5\ \text{cm}^{-3}$ and a relative velocity of $v=10$ km/s. The angle between the BH rotation axis and the line of sight was fixed to $\pi/2$. The dot-doshed line is the corresponding border for $\pi/8$. Inside the gray region, the electrons are relativistic at the conversion point. We also fixed $c_\gamma=1$ in Eq. \ref{eq: axion photon coupling}.}
    \label{fig:LOFAR plots}
\end{figure*}

In this section, we study the prospects of detecting the signals studied in the previous section with the radio telescope. 

The frequency $\nu$ of the photons emitted from the conversion is related to the mass of the non-relativistic axions in the cloud:
\begin{equation}
    \nu=\frac{m_a}{2\pi}\simeq 0.24\ \parfrac{m_a}{10^{-9} \text{ eV}} \text{ MHz} \, .
\end{equation}
However, the requirement that the electrons in the plasma are non-relativistic at the conversion radius (see Eq. \ref{eq: plasma temperature}) requires, for the hot ionized medium, that  $m_a\lesssim 10^{-9}$ eV and so a maximal photon frequency of approximately $0.2$ MHz which is hard to measure with Earth-based telescopes because of the ionosphere screening at 10 MHz, although space-based radio telescopes might allow to go beyond this boundary \cite{Cecconi_2014,Cecconi_2018o}. Nonetheless, for the warm ionized media the bound is weaker, $\nu_\text{max} <100$ MHz, thus leaving open the possibility of detection with radio telescopes.
In particular, at these low-radio frequencies the radio telescope LOFAR is the one with the best sensitivities and we therefore proceed with a dedicated study of detectability with LOFAR \cite{vanHaarlem:2013dsa,Lofarlink}.

The relevant quantity for radio observations is the flux density that is related to the luminosity emitted per solid angle computed in Sec. \ref{sec:Resonant conversion wo gravitational potential} by 
\begin{equation}
    S= \frac{1}{d^2\Delta\nu}\frac{dP}{d\Omega}
\end{equation}
where $d$ is the distance to the source, and $\Delta \nu$ is the typical width of the signal which is related to the variance in the effective velocity of the axions in the cloud $\Delta \nu/\nu \sim \left<v_\text{cl}^2 \right> = \alpha/2$ \cite{Rosa_2018} since the relative velocity between the BH and the medium is much smaller than $\alpha/2$.

Regarding the typical distance to the closest PBH  ($d$), from Fig. \ref{fig:Density plot}, we concluded that photon emission is mostly efficient for PBH with subsolar masses, between $10^{-6}\ -\ 10^{-1}\ M_{\odot}$ depending on the medium.  In this range of masses, the PBH cannot account for all the dark matter but can still account for a fraction $f_\text{PBH}=1-10\%$ of the total \cite{Green_2021}. By fixing $f_\text{PBH}$ we can then estimate the typical distance to the closest PBH to be
\begin{eqnarray} \label{eq: distance}
    d&=&\frac{1}{n_\text{PBH}^{1/3}}=\left(\frac{M_\text{BH}}{f_\text{PBH}\ \rho_\text{DM}}\right)^{1/3}\\
    &=& 0.33\ \parfrac{0.1}{f_\text{PBH}}^{1/3}\left(\frac{M_\text{BH}}{10^{-5}M_\odot}\right)^{1/3} \text{ly}
\end{eqnarray}
where $n_\text{PBH}$ is the number density of PBH. We fixed the dark matter density to the local value $\rho_\text{DM}=0.36$ GeV/$\text{cm}^3$ \cite{sofue2020rotation} and assumed the PBHs to be roughly uniformly distributed within this region. It is then likely that the closest PBH is within our Local Bubble which is characterized by low densities and large ionization fractions \cite{Alves2018}. 
Then, using the previous equation and the luminosity calculated in  Eq. \ref{eq: Luminosity} we find that, for some benchmark values of the parameters,  the flux density $S$ in Janksys that is expected from the axion-photon conversion is
\begin{align}
    S&=2.54\times 10^{10}\ \tilde{a}\parfrac{\tilde{n}_\text{crit}}{N} \,\parfrac{f_\text{PBH}}{0.1}^{2/3}\parfrac{100 \,\text{km/sec}}{v}^{4} \nonumber\\
    & \times \parfrac{n_\infty}{10^{-3}\, \text{cm}^{-3}}^{4/3}
    \parfrac{M_\text{BH}}{M_\odot}^{16/3}\parfrac{m_a}{10^{-9}\text{ eV}}^{13/3}\ \text{Jy}.
    \label{eq: S system}
\end{align}
Let us now compare this value with the sensitivity reach of LOFAR that is currently formed by an array of 52 stations spread across northern Europe. Each station has low and high-band antennas; the low-band ones are sensitive to frequencies in the range $10\ -\ 90$ MHz that we are interested in.
The minimum energy flux that a given station can detect is given by the radiometer equation by \cite{vanHaarlem:2013dsa,Lofarlink}
\begin{equation}
S_\text{station}=\frac{2k_BT_\text{sys}}{A_\text{eff}\sqrt{\Delta\nu\Delta t}}
\end{equation}
where $k_B$ is the Boltzmann constant, $T_\text{sys}$ is the noise temperature of the system, $A_\text{eff}$ is the effective area, and $\Delta t$ the observation time. For LOFAR, we take the fiducial values $T_\text{sys}=60 (\lambda/(1\ \text{m}))^{2.55}$ Kelvin, and $A_\text{eff}=48 \lambda^2/3$, where $\lambda$ is the wavelength of the signal \cite{vanHaarlem:2013dsa}. In the case of an array of $N=52$ stations, the sensitivity is
\begin{eqnarray}
S_\text{array}=\frac{2k_BT_\text{sys}}{A_\text{eff}\sqrt{2N(N-1)\Delta\nu\Delta t}} 
\end{eqnarray}
which in terms of our model parameters corresponds  to 
\begin{eqnarray}
S_\text{array}=23\parfrac{M_\odot}{M_\text{BH}}\parfrac{10^{-9}\text{ eV}}{m_a}^{2.05}\parfrac{8 \text{ h}}{\Delta t}^{1/2} \text{mJy} \, .\,\,
      \label{eq: S array}
\end{eqnarray}

We then perform a sensitivity analysis where we look for the parameter space where the flux density emitted by the axion-photon conversion in Eq. \ref{eq: S system} is larger than the sensitivity levels of LOFAR in Eq. \ref{eq: S array}, with typical observation time of 8 hours. 

The results are shown in Fig. \ref{fig:LOFAR plots}.
The red region is the range of PBH and QCD axion masses where the signal can in principle be detected at LOFAR with a dedicated analysis. In the innermost part of the red region, the flux density is rather large and the signal can be detected with high significance. In each plot, the red region is enclosed by the blue band which corresponds to the range of parameters where the emitted photons lie in the frequency band of the LOFAR low-band array while satisfying the superradiant constraints as in Fig. \ref{fig:Density plot}. Inside the gray region, the electrons in the plasma are relativistic at the conversion point and so outside the range of validity of the expressions we used.

In the analysis, we fixed the velocity parameter to $v=10$ km/s and the PBH fraction to $f_\text{PBH}=0.1$. We have also chosen an angle between the line of sight and the rotation axis of the BH of $\pi/2$ but showed the results for another representative value of $\pi/8$ (dot-dashed curve).
We then considered two different values of the medium density, $n_\infty=0.5 \text{ cm}^{-3}$ and $0.2 \text{ cm}^{-3}$, typical of the warm ionized medium, and two different values of the initial PBH spin, $\tilde{a}=0.1$ and close to $1$. Note that by decreasing $\tilde{a}$ the superradiant condition $\alpha < \tilde{a}/2$ becomes stronger and therefore the allowed parameter space is smaller.
In particular, for $\tilde{a}=0.01$ the parameter space is too narrow and there is no detectable region. 

The red region in Fig. \ref{fig:LOFAR plots} corresponds to PBH masses around $10^{-5}-10^{-4}\ M_\odot$. Coincidently, this range of PBH masses is within the range that can explain the ultra-short-timescale microlensing events measured by the OGLE experiment (see Fig. 8 of \cite{Niikura:2019kqi}).

\section{Discussion and Outlook}
\label{sec:Outlook}

The main outcome of this work was to show that the QCD axion can illuminate the vicinity of PBH with subsolar masses and that if the latter amounts to $1-10\%$ of the total dark matter abundance the resulting signals could be detected at radio telescopes such as LOFAR. 

The mechanism relied on two main ingredients, superradiance and the accretion onto the PBH. On one hand, the superradiant instability dynamically creates a dense cloud of axions around the PBH and therefore our results do not assume an initial QCD axion profile.
On the other hand, the accretion of the interstellar medium ensures the presence of a critical radius, where the QCD axion mass and the effective plasma mass are equal, and therefore where the resonant conversion can occur. As explained in section \ref{sec:Interstellar medium and accretion}, we used a model of Bondi accretion and assumed that the plasma behaves non-relativistically in the relevant region for the axion-photon conversion. This condition is fulfilled far away from the Schwarzschild radius, which allowed us to also neglect gravitational corrections.

To compute the luminosity that results from the conversion, we considered in section \ref{sec:Resonant conversion wo gravitational potential} a Klein-Gordon-like equation for the component of the photon field parallel to the external magnetic field, which is the one that interacts with the QCD axion. For simplicity, we assumed this magnetic field to be perpendicular to the line of sight in the relevant region for the conversion, and neglected ray-bending effects. We also neglected self-interactions and backreaction in the QCD axion cloud. The former is justified by the fact that we restricted to subsolar PBHs with masses such that self-interactions are only relevant above the QCD axion line \cite{Baryakhtar:2020gao};  the latter because of the restriction to non-relativistic plasmas at the conversion point that enforces that
the interesting magnetically stimulated regime that we discuss in Appendix \ref{app:Stimulation emission}, where the produced magnetic fields are larger than the original ones, does not happen. 

In solving the photon equation for finding the luminosity of the system we solved the second-order photon equation instead of reducing it to a first-order one, which is common practice in the literature. We found that, under two simplifying assumptions that hold in the most efficient part of the allowed parameter space, i.e. when the critical point and the most populated part of the superradiant cloud are not far from each other, most photons are produced around the critical radius. 
These simplifications, also allowed us to arrive at an analytical expression for the luminosity in Eq. \ref{eq: Luminosity} that is similar to that in the axion-photon conversion literature. 

The resulting luminosity of the process was shown in Fig. \ref{fig:Density plot} for different interstellar media. In the conservative region of the parameter space, we found luminosities as high as $10^{12}$~W for an accretion from the warm ionized medium. These high luminosities confirmed our hypothesis: even if the magnetic field in the ISM is very weak when compared with the magnetic fields around neutron stars or magnetars, this is highly compensated by the huge amount of QCD axions generated by superradiance. 

Finally, in section \ref{sec:Fate of the conversion} we studied the possibility of directly detecting the emitted electromagnetic signals with existing radio telescopes. We considered the low band antenna of the LOFAR telescope, which is the one with the best sensitivity in the conservative region of our parameter space, and found that PBHs and QCD axions in the mass range between $10^{-5}-10^{-4}\, M_\odot$ and $4\times 10^{-7}-4\times 10^{-8}$, respectively, could potentially be detected (see Fig \ref{fig:LOFAR plots}). Interestingly, this region coincides with the range of PBH masses used in \cite{Niikura:2019kqi} to explain some microlensing events observed by the OGLE experiment. This is a clear motivation to look for further evidence in the LOFAR data.

This work has left a few interesting avenues that we believe will be interesting to pursue in more detail in the future.
We focused on the possibility of directly detecting the light coming from the closest PBH but it would be interesting to study the stochastic signal that would be produced by the cosmological PBH population. 
It would also be interesting to extend the results to accretion models that account for angular momentum as the advection-dominated accretion flow \cite{Mahadevan_1997}, and to explore the possibility of large magnetic fields via the Blandford instability \cite{Blandford:1977ds}.  
On the other hand, although we have restricted to cases where the plasma is non-relativistic around the conversion radius, we do not expect significant changes in the plasma mass in the relativistic case and therefore the prospects that we find in this work likely extend to the relativistic region. Moreover, this would allow signals at larger frequencies that could be probed at other radio telescopes. 
Finally, the methods we developed are directly extensible to axion-like particles (ALPs), which opens up the parameter space since the mass and the coupling are not correlated.

\section{Acknowledgements}
We would like to thank Erik Engstedt for participation in the early stages of this project and M.C. David Marsh and João Rosa for discussions.
R.Z.F. acknowledges the financial support provided through national funds by FCT - Fundação para a Ciência e Tecnologia, I.P., reference 2022.03283.CEECIND as well as the FCT projects CERN/FIS-PAR/0027/2021, UIDB/04564/2020 and UIDP/04564/2020,  with DOI identifiers 10.54499/CERN/FIS-PAR/0027/2021, 10.54499/UIDB/04564/2020 and 10.54499/UIDP/04564/2020, respectively.  
Á.G.M. has been supported by the Secretariat for
Universities and Research of the Ministry of Business and Knowledge of the Government of
Catalonia and the European Social Fund. Á.G.M. acknowledges the
support from the Departament de Recerca i Universitats from Generalitat de Catalunya to the Grup
de Recerca ‘Grup de Física Teòrica UAB/IFAE’ (Codi: 2021 SGR 00649). IFAE is partially
funded by the CERCA program of the Generalitat de Catalunya.

\onecolumngrid

\appendix

\section{Components of the interstellar medium}
\label{app:Media}

\begin{center}
 Table 1: Components of the interstellar medium and reference values for some of their properties. \\ Adapted from \cite{McKee_1977,Draine_2010,Ferriere_2001,Ioka:2016bil}. 
\begin{equation} \label{tab: Components of the interstellar medium}
\begin{aligned} 
&\begin{array}{|c|c|c|c|c|}
\hline \text { Component } & \begin{array}{c}
    \text { Temperature } \\
    \mathbf{( K )} \text{  }
\end{array} & \begin{array}{c}
    \text {Density} \\
    \text {(cm}^{-3}\text{)   } 
\end{array} & \begin{array}{c}
    \text {Sound speed} 	\\
    \text {(km/sec) }
\end{array} &  \begin{array}{c} \text{Ionized fraction}  \\ \text{} x_e\end{array}  \\
\hline \text { Molecular clouds }  &  10-50  & 10^{2}-10^{6} & 0.6 & \begin{array}{c}
    10^{-7}-10^{-4}
\end{array} \\
\hline \begin{array}{c}
    \text { Cold neutral } \\
    \text { medium (CNM) }
\end{array} & \begin{array}{c}
      100
\end{array}  & 30 & 10& 10^{-3} \\
\hline \begin{array}{c}
    \text { Warm neutral } \\
    \text { medium (WNM) }
\end{array} &   5000  & 0.2-0.5 &10 & \multicolumn{1}{|c|} {0.1} \\
\hline \begin{array}{c}
    \text { Warm ionized } \\
    \text { medium (WIM) }
\end{array}  &   10^{4} & 0.2-0.5 & 10 &  1  \\
\hline \begin{array}{c}
    \text { Hot ionized } \\
    \text { medium (HIM) }
\end{array}  & 10^{6} & 10^{-3}-10^{-2} & \begin{array}{c}
     100
\end{array} & \begin{array}{l}
    1
\end{array} \\
\hline
\end{array}
\end{aligned}
\end{equation}
\end{center}

\section{Method of variation of parameters \label{app: Method of variation of parameters}}

In this appendix, we provide the details that allowed us to arrive at the asymptotic solution for the field $A(t,\vec{r})$ in Eq. \ref{eq: field} using the method of variation of parameters. Starting from Eq. \ref{eq: Photon equation}
\begin{equation} 
   \left(-m_a^2-\nabla^2+\omega_{p}^2(r)\right)A(\Vec{r})=f(\Vec{r})
   \label{2} \, ,
\end{equation}
and after decomposing the source $f(\Vec{r})$ and $A(\Vec{r})$ in spherical harmonics, we arrive at:
\begin{equation}
    \partial_r^2(rA_{lm}(r)) + \left[m_a^2-\omega_{p}^2(r)-\frac{l(l+1)}{r^2}\right](rA_{lm}(r))=-r f_{lm}(r)
    \label{eq: u eq}
\end{equation}
where $A_{lm}(r)$, $f_{lm}(r)$ are respectively the projections of the field and the source over the harmonics.

Since $\omega_{p,\infty}$ is a non-zero constant, the homogeneous solutions of Eq. \eqref{eq: u eq} are $F_l(r),\ G_l(r)$~\footnote{Eq. \eqref{eq: u eq} does not depend on the angular number $m$, so for each mode the homogeneous solutions depend only on $l$.}  such that $ F_l(r\to\infty) \to \cos(\theta_l(k_\infty r))$ and $G_l(r\to\infty) \to \sin(\theta_l(k_\infty r))$, with $k_\infty^2=m_a^2-\omega_{p,\infty}^2$, where $\partial_r\theta_l\to k_\infty$ when $r\to\infty$. 
For convenience, in the following, we choose  $F_l$ and $H_l(r)=F_l(r)+iG_l(r)$  as the pair of independent solutions. Then, the particular solution of \ref{2} can be written as
\begin{equation}
    A_{lm}(r)=\frac{F_l(r)}{r}\int_r^\infty\frac{r' f_{lm}(r')}{W(F_l,H_l)}H_l(r')\ dr'\ +\ \frac{H_l(r)}{r}\int_0^r\frac{r' f_{lm}(r')}{W(F_l,H_l)}F_l(r')\ dr'
    \label{eq: u exp}
\end{equation}
where $W(F_l,H_l)=F_l(r)H'_l(r)-F'_l(r)H_l(r)$ is the Wronskian operator. The constant limits of integration are arbitrary; here they have been chosen to fix initial conditions. For $r\to\infty$, we fixed the solution to behave  like $H_l(r)/r=e^{i\theta_l(r)}/r$, and for $r\sim0$ as $F_l(r)/r$. The choice of  $F_l$ rather than $G_l$ is related to the fact that, as can be seen in the next appendix, $\lim_{r\to 0}F_l(r)=0$; this is a sensible initial condition since in our setup the plasma mass is very large at small radii and the conversion into photons becomes kinematically forbidden.

In  general,  the Wronskian for two solutions $\phi_1,\ \phi_2$ of a second order differential homogeneous equation without first-order term $\phi_{1,2}''(r)+p(r)\phi_{1,2}(r)=0$, like Eq. \eqref{eq: u eq}, is constant. Using this together with the fact that the Wronskian is a multilinear operator, we can find its value to be:
\begin{equation}
    W(F_l,H_l)=iW(F_l,G_l)\to iW(\cos\theta_l(k_\infty r),\sin\theta_l(k_\infty r))=ik_\infty\, .
\end{equation}

In the end, we are interested in the luminosity at infinity, so we substitute the asymptotic expressions in Eq. \eqref{eq: u exp} and find
\begin{align}
    A_{lm}(r\to \infty)&= \frac{e^{i\theta_l(k_\infty r)}}{ik_\infty r}\int_0^\infty r' f_{lm}(r')F_l(r')\ dr'\,.
\end{align}
The asymptotic expression for the photon field is $A(\Vec{r})= \sum_{lm}Y_{lm}(\hat{r}) A_{lm}$.  In our case, the source only has non-vanishing contributions for $l=m=1$, so the asymptotic expression further simplifies to 
\begin{equation}
    A(\Vec{r})\to Y_{11}(\hat{r})\frac{e^{i\theta_l(k_\infty r)}}{ik_\infty r}\int_0^\infty r' f_{lm}(r')F_l(r')\ dr'
    \label{field}
\end{equation}
with radial derivative $\partial_rA(\Vec{r})=ik_\infty A(\vec{r})$,
where we used that $\partial_r\theta_l(k r)\to k$ and neglected terms proportional to $1/r^2$.

\section{Solution of the homogeneous equation \label{app: Solution of the homogeneous equation}}

In appendix \ref{app: Method of variation of parameters} we derived the general formula for the luminosity in terms of a convolution of the homogenous solution of the differential equation with the source term. 
Now we proceed with the derivation of the homogeneous solution $F_l(r)$ of Eq. \ref{eq: u eq}. We start by rewriting the differential equation in terms of the adimensional parameter $\rho=m_a r$ as
\begin{equation}
 \partial_\rho^2(F(\rho)) + \tilde{k}(\rho)^2F(\rho)=0
    \label{eq: Homogeneous eq}
\end{equation}
where $\tilde{k}(\rho)^2= 1-\left(\rho_\text{crit}/\rho \right)^{3/2}$ is the effective momentum in units of $m_a$, we neglected the small angular momentum contribution (and for that reason we drop the subindex $l$) and defined $\rho_\text{crit} \equiv m_a r_\text{crit}$. 

We can now use the WKB method to find the solution of the equation of motion for both large $\rho > \rho_\text{crit}$ and small $\rho < \rho_\text{crit}$. In the intermediate region, $\rho \sim \rho_\text{crit}$, where the WKB approximation does not hold, we will instead Taylor expand the effective momentum $\tilde{k}$ and look for exact solutions that we then match to the WKB regimes for large and small $\rho$.

\paragraph{$\rho > \rho_\text{crit}$:}
When $\rho > \rho_\text{crit}$
the general WKB solution 
is given by
\begin{equation} \label{eq: cos branch}
    F(\rho)= \sqrt{\frac{1}{\tilde{k}(\rho)}}\left(a_1\cos\left[ \int_{\rho_\text{crit}}^\rho \tilde{k}(\rho')d\rho'-\pi/4\right]+a_2\sin\left[ \int_{\rho_\text{crit}}^\rho \tilde{k}(\rho')d\rho'-\pi/4\right]\right), \, \qquad \rho > \rho_\text{crit}.
\end{equation}
The lower limit of integration and the $\pi/4$ phase are taken for convenience. In app. \ref{app: Method of variation of parameters} we required $F(\rho)\to\cos(\theta(kr))$ as $r\to\infty$, so we need to set $a_2=0$ and $a_1=\sqrt{\tilde{k}_\infty}$.\footnote{ Actually, $F$ is an exact solution of \ref{eq: Homogeneous eq} for  $\rho>\rho_b$ where $\rho_b$ is the value of $\rho$ at the Bondi radius where the momentum approaches a constant, $\tilde{k} \to k_\infty/m_a$ .}

\paragraph{$|\rho -\rho_\crit | \ll \tilde{L}^2/(2\pi)$:}
The WKB approximation breaks down around the resonant point $\rho=\rho_\text{crit}$. To find a solution in this regime we Taylor expand $\tilde{k}(\rho)$ around $\rho_\text{crit}$ to first order, 
\begin{eqnarray} \label{eq: Taylor expansion}
   \tilde{k}(\rho)^2\sim\frac{2\pi}{\tilde{L}^2}(\rho-\rho_\text{crit}) 
\end{eqnarray}
where $\tilde{L}=m_a\sqrt{\pi/\omega_p'(\rho_\text{crit})}=\sqrt{4\pi\rho_\text{crit}/3}$ is the width of the critical region in units of $m_a$ (the derivative is taken with respect to $r$). We then obtain the differential equation:
\begin{align}
    \partial_\rho^2(F(\rho)) + \frac{2\pi}{\tilde{L}^2}(\rho-\rho_\text{crit})F(\rho)=0
\end{align}
whose solutions are Airy functions
\begin{align}
F(\rho)=b_1\ \text{AiryAi}\left[\left(\frac{\tilde{L}^2}{2\pi}\right)^{-1/3}(\rho_\text{crit}-\rho)\right]+b_2\ \text{AiryBi}\left[\left(\frac{\tilde{L}^2}{2\pi}\right)^{-1/3}(\rho_\text{crit}-\rho)\right], \, \qquad|\rho -\rho_\text{crit}| \ll \tilde{L}^2/(2\pi) 
    \label{eq: Airy branch} 
\end{align}
and $b_1, b_2$ are integration constants.

To find these constants we match the solutions for $F$ in the overlapping region of validity of both branches which is when $\tilde{L}^2/(2\pi)\gg\rho-\rho_\crit \gg \left(\tilde{L}^2/(2\pi)\right)^{1/3}$.
Namely, the Taylor expansion of $\tilde{k}(\rho)^2$ in \ref{eq: cos branch} yields
\begin{align}
F(\rho)\sim\sqrt{\tilde{k}_\infty}\left(\frac{\tilde{L}^2}{2\pi}\right)^{1/4}(\rho-\rho_\text{crit})^{-1/4}\cos\left[ \frac{2\sqrt{2\pi}}{3\tilde{L}}(\rho_\text{crit}-\rho)^{3/2}-\pi/4\right] 
    \label{eq: asympt behaviour 1}\,,
\end{align}
which we then match to the asymptotic expansion of Eq. \ref{eq: Airy branch} for $\rho\gg\rho_\text{crit}$
\begin{align}
    F(\rho)\simeq {\sqrt{\pi}}\left(\frac{\tilde{L}^2}{2\pi}\right)^{1/12}(\rho-\rho_\text{crit})^{-1/4} \left(b_1 \cos\left[ \frac{2\sqrt{2\pi}}{3\tilde{L}}(\rho_\text{crit}-\rho)^{3/2}-\pi/4\right]
    -b_2 \sin\left[ \frac{2\sqrt{2\pi}}{3\tilde{L}}(\rho_\text{crit}-\rho)^{3/2}-\pi/4\right] \right)
\end{align}
to find $b_2=0$ and $b_1=\sqrt{\pi\tilde{k}_\infty}(\tilde{L}^2/(2\pi))^{1/6}$.

\paragraph{$\rho < \rho_\text{crit}$: }
Finally, we look at the region where $\rho<\rho_\text{crit}$. The general WKB solution is 
\begin{equation}
    F(\rho)=\frac{1}{\sqrt{|\tilde{k}(\rho)|}}\left(c_1\ \text{Exp}\left[\int_{0}^\rho |\tilde{k}(\rho')|d\rho'\right]+c_2\ \text{Exp}\left[-\int_{0}^\rho |\tilde{k}(\rho')|d\rho'\right]\right), \qquad \rho < \rho_\text{crit}
    \label{eq: WKB 2} \, .
\end{equation}
The integration limits were chosen for convenience, and $c_1,c_2$ are integration constants that we again fix by matching Eqs. \ref{eq: Airy branch} and \ref{eq: WKB 2} in the common region of validity, i.e. $\tilde{L}^2/(2\pi)\gg \rho_\crit - \rho \gg \left(\tilde{L}^2/(2\pi)\right)^{1/3}$. Namely, we match the expansion of Eq. \ref{eq: Airy branch} for $\rho_\crit - \rho \gg \left(\tilde{L}^2/(2\pi)\right)^{1/3}$
\begin{equation}
    F(\rho < \rho_\text{crit})\rightarrow \frac{\sqrt{\tilde{k}_\infty}}{2}\ \left(\frac{\tilde{L}^2}{2\pi}\right)^{1/4}(\rho_\text{crit}-\rho)^{-1/4}\text{Exp}\left[\frac{-2\sqrt{2\pi}}{3\tilde{L}}(\rho_\text{crit}-\rho)^{3/2}\right]
    \label{eq: asympt 2}\,,
\end{equation}
where we used the values of $b_1,b_2$ found before, with the Taylor expansion of Eq. \ref{eq: WKB 2} close to the critical radius, $|\tilde{k}(\rho)|\sim\left(\frac{2\pi}{L^2}(\rho_\text{crit}-\rho)\right)^{1/2}$,
\begin{align} \label{eq: F small rho}
    F( \rho < \rho_\text{crit}) \sim c_1 \left(\frac{\tilde{L}^2}{2\pi}\right)^{1/4}(\rho_\text{crit}-\rho)^{-1/4}C^{-1}\text{Exp}\left[\frac{-2\sqrt{2\pi}}{3\tilde{L}}(\rho_\text{crit}-\rho)^{3/2}\right]\, ,
\end{align}
where $c_2$ was fixed to zero and
\begin{equation}
    C\equiv\text{Exp}\left[-\int_{0}^{\rho_\text{crit}} |\tilde{k}(\rho')|d\rho'\right]=\text{Exp}\left[-\frac{2\sqrt{\pi}\Gamma(7/6)}{\Gamma(5/3)}\rho_\text{crit}\right] \, ,
\end{equation}
and find $c_1=C\sqrt{\tilde{k}_\infty}/2$. Note that $\text{lim}_{\rho\to0}F(\rho)=0$, consistently to what we imposed in Appendix \ref{app: Method of variation of parameters}.~\footnote{The WKB approximation  breaks down when $\rho\sim0$. In that limit, the exact solution, which can be found exactly using $\tilde{k}^2\sim \left(\rho_\text{crit}/\rho\right)^{3/2}$, is a Bessel function that also fulfills the requirements of $F_{\rho}$.}

We can now put all the solutions together to find the complete solution for $F(\rho)$
\begin{align}
    F(\rho)=
    \begin{dcases}
       \frac{C}{2}\sqrt{\frac{\tilde{k}_\infty}{|\tilde{k}(\rho)|}}\ \text{Exp}\left[\int_{0}^\rho |\tilde{k}(\rho')|d\rho'\right]\hspace{0.1cm} &,\, \rho < \rho_\text{crit}\\
        \sqrt{\pi\tilde{k}_\infty}(2\pi)^{-1/6}\tilde{L}^{1/3}\text{AiryAi}\left((2\pi)^{1/3}\tilde{L}^{-2/3}(\rho_\text{crit}-\rho)\right)\hspace{0.1cm} &, \, |\rho -\rho_\text{crit}| \ll \tilde{L}^2/(2\pi) \\
       \sqrt{\frac{\tilde{k}_\infty}{\tilde{k}(\rho)}}\cos\left[ \int_{\rho_\text{crit}}^\rho \tilde{k}(\rho')d\rho'-\pi/4\right]\hspace{0.1cm} &, \,\rho > \rho_\text{crit}
    \end{dcases}
    \label{eq: Homogeneous sol}
\end{align}

\section{Analytical computation of the luminosity}
\label{app:Analytics}

In this appendix, we  derive a simplified formula for the photon luminosity that results from the axion to photon conversion and elaborate under which conditions the formula applies. From Eqs. \ref{eq: field} and  \ref{eq: Lum def}, the expression for the luminosity emitted per solid angle is 
\begin{align}
    \left<\frac{dP}{d\Omega}\right>=\left|Y_{11}(\hat{r})\right|^2\ \frac{m_a^2}{k_\infty}g_{a\gamma\gamma}^2N\left|\int B(r')R_{21}(r')F(r')r'dr'\right|^2
    \label{eq: app Luminosity} \,.
\end{align}
 For $\rho<\rho_\text{crit}$ (recall that $\rho=m_a r$) the homogenous solution $F(r)$  is exponentially suppressed if $\rho_\text{crit}\gg 1$, so we can neglect this contribution to the integral. Similarly, the region $\rho>\rho_\text{crit}$ can be neglected due to the highly oscillatory behavior of $F(r)$. Therefore, the main contribution to the luminosity comes from the region $|\rho -\rho_\text{crit}| \ll \tilde{L}^2/(2\pi)$, i.e. from the region around the resonance $m_a \simeq \omega_p(r_\text{crit})$.~\footnote{Numerically we verified that for $\rho_\crit \gtrsim 5$ and $\alpha$ within the range of interest, the region around the resonance dominates over the others.}

However, we can further simplify the integral by noting, as we will do next, that in the cases of interest most of its support is in an even smaller window of size proportional to $\tilde{L}^{2/3}$ which is parametrically smaller than the typical scale of variability of $g(\rho)\equiv B
(\rho) R_{21}(\rho)\rho$ around $\rho_\crit$ which is 
\begin{equation} \label{eq: region where B R are constant}
    \left|\frac{g'(\rho_\text{crit})(\rho-\rho_\text{crit})}{g(\rho_\text{crit})}\right|<1 \iff |\rho- \rho_\text{crit} |<\frac{4\rho_\text{crit}}{|3-2\alpha\rho_\text{crit}|} \,.
\end{equation}
To verify that, we will approximate $g(\rho)\sim const.$ and then check by inspection under what conditions the remaining integral has most of its support in a region smaller than Eq. \ref{eq: region where B R are constant}. Following these steps, we find that
\begin{align}
    \int_{\rho_0}^{\rho_f} F(\rho')d\rho'=\sqrt{\frac{\tilde{k}_{\infty}}{2}}\tilde{L}\Bigg[1-\frac{1}{2\sqrt{\pi}(c\Delta_0)^{3/4}}e^{-2(c\Delta_0)^{3/2}/3}+\frac{1}{\sqrt{\pi}(c\Delta_f)^{3/4}}\sin\left(\frac{\pi}{4}-\frac{2}{3}(c\Delta_f)^{3/2}\right)\Bigg]
    \label{eq: Analytical integral}
\end{align}
 where $c=(3\rho_\text{crit}^{2}/2)^{1/3}$, $\Delta_0=(1-\rho_0/\rho_\text{crit})$, $\Delta_f=(\rho_f/\rho_\text{crit}-1)$. Whenever $c \Delta_0, c\Delta_f \gg 1$ or, correspondingly, when the integration window is $|\rho_f-\rho_0| \gg \rho_\crit^{1/3}= (3\tilde{L}^{2}/(4\pi))^{1/3}$, the integral becomes insensitive to the integration limits and simplifies to
\begin{align}
    \int_{\rho_0}^{\rho_f} F(\rho')d\rho'=\sqrt{\frac{\tilde{k}_{\infty}}{2}}\tilde{L} \, .
    \label{eq: Analytical integral 2}
\end{align}
Therefore, the support of the integral is inside the region in Eq. \ref{eq: region where B R are constant} where the function $g(\rho)$ is approximately constant, if
\begin{equation} \label{eq: constant B and cloud profile approximation}
    \left|\frac{4\rho_\text{crit}}{3-2\alpha\rho_\text{crit}}\right|\gtrsim \rho_\text{crit}^{1/3} \, .
\end{equation}
If this condition, together with $\rho_\text{crit}>1$, is fulfilled, then we can further simplify Eq. \ref{eq: app Luminosity} and find
\begin{align}
    \left<\frac{dP}{d\Omega}\right>&=\left|Y_{11}(\hat{r})\right|^2\ \frac{m_a^2}{k_\infty}g_{a\gamma\gamma}^2B_\text{crit}^2N R_{21}(r_\text{crit})^2r_\text{crit}^2\left|\int F(r')dr'\right|^2  \simeq m_a\ n_\text{crit}\ r_\text{crit}^2\frac{(L\ g_{a\gamma\gamma}\ B_\text{crit})^2}{2}
\end{align}
where he have defined $n_\crit= N R_{21}(r_\text{crit})^2$ and used Eq. \ref{eq: Analytical integral 2}. Interestingly,
this last term is reminiscent of the standard expression for the axion-photon conversion probability \cite{Raffelt_1987}. This might not be surprising as we have concluded that the magnetic field and the axion cloud are approximately constant in the conversion region and therefore the conversion becomes closer to the more standard processes studied in the literature.

Note that the fact that the luminosity at infinity does not depend on the momentum at infinity can be counter-intuitive. However, one should take into account that the energy released at conversion is fixed: the evolution in the plasma mass afterward does not change the flux. Taking the particle perspective, a change in the photon mass can be understood as a change in the number of photons. Conservation of energy implies that the energy flux is the same.

\section{Magnetic stimulation and cloud quenching}
\label{app:Stimulation emission}

The photons converted from the axion cloud will themselves generate an electric and magnetic field. When this new induced magnetic field $B_\text{ind}$ becomes comparable
    \begin{align}
 \langle|B_\text{ind}(r)|^2\rangle \sim {B^2}(r)
\end{align}
with the background interstellar magnetic field $B$ that generated the axion-to-photon conversion in the first place (the brackets denote a time-average) the system is expected to enter a \textit{magnetically stimulated} stage where the conversion rate could grow very fast, in a snowball effect, analogously to other stimulation effects that have been studied in the literature (e.g. \cite{Rosa_2018,Ikeda_2019,Blas_2020b}) even though here the process is initially linear, one axion converts to one photon. 

To understand the region of parameters where this might happen we  estimate the induced magnetic field as 
\begin{eqnarray}
    \langle |B_\text{ind}(r)|^2\rangle=\left|\frac{2}{r}\partial_r\left(rA\right)\right|^2
    \sim \frac{a_0^{-3} \tilde{n}_\text{crit}}{2} \frac{k(r)^2}{k_\infty^2}  \left( g_{a\gamma\gamma}B(r) L \right)^2\,
\end{eqnarray}
where we have used the fact that the conversion is highly dominated by the dynamics around the critical radius, and therefore the field $A(t,\vec{r})$ will behave like Eq. \ref{eq: field} shortly after that point, and assumed Eq. \ref{eq: Luminosity} to be satisfied.

The maximum value of the previous expression takes place at the maximum of $k(x)^2/x^2$ is, with $x=\rho/\rho_\text{crit}$, and turns out to be $\simeq 0.2$, at $\rho \simeq1.45\rho_\text{crit}$. Then, to be sure that the stimulated resonance does not occur, we impose
\begin{align}
\text{max}\left[\frac{\langle|B_\text{ind}|^2\rangle}{B^2} \right]=\frac{0.2m_a^5}{k_\infty}\ \frac{(g_{a\gamma\gamma} L)^2}{2}\tilde{n}_\text{crit}\big|_{\theta=\pi/2}<1
    \label{eq: Magnetic stim bound}
\end{align}
where we chose an angle of $\pi/2$ since $|Y_{11}|^2$ is maximized in that direction.
Note that if the radius at which the ratio is maximized lies outside the relevant interval for conversion, the axion-photon conversion might not be affected by the magnitude of $|B_\text{ind}|^2$. For this reason, Eq. \ref{eq: Magnetic stim bound} is not an exact characterization of the magnetic stimulation, but rather a conservative bound to ensure that such stimulation does not occur.

Although this regime might be the one leading to the largest observable signatures, as it might lead to an explosion of the cloud into photons, to have a robust understanding of what happens beyond the threshold of magnetic stimulation we would need to backreact the effect of the cloud on the dynamics of the accretion. Moreover, this constraint is weaker than that coming from imposing the electrons to be non-relativistic at the conversion radius.  Therefore, we conservatively discard it in the main text.

\twocolumngrid

\bibliography{refs}

\end{document}